\documentclass[aps,prl,twocolumn,groupedaddress,showkeys,showpacs]{revtex4}

\usepackage{multirow,amssymb,amsbsy,amsmath,bm}
\usepackage{epsfig}

\newcommand {\mofe} {\{$\textrm{Mo}_{72}\textrm{Fe}_{30}$\}}

\begin{document}

\title{Competing Spin Phases in Geometrically Frustrated Magnetic Molecules}

\author{Christian Schr\"oder}
\email{schroder@ameslab.gov}
\affiliation{Department of Electrical Engineering and Computer
  Science, University of Applied Sciences Bielefeld, D-33602 Bielefeld, Germany and Ames Laboratory, Ames, Iowa 50011, USA}

\author{Hiroyuki Nojiri}
\affiliation{Department of Physics, Okayama University, Tsushimanaka 3-1-1, Okayama, 700-8530 Japan}

\author{J\"urgen Schnack}
\author{Peter Hage}
\affiliation{Universit\"at Osnabr\"uck, Fachbereich Physik,
D-49069 Osnabr\"uck, Germany}

\author{Marshall Luban}
\author{Paul K\"ogerler}
\affiliation{Ames Laboratory \& Department of Physics and Astronomy,
Iowa State University, Ames, Iowa 50011, USA}

\date{\today}

\begin{abstract}
We have identified a class of zero-dimensional classical and quantum Heisenberg
spin systems exhibiting anomalous behavior in an external magnetic field $B$ similar to that found for the geometrically frustrated Kagom\'e lattice of classical spins. 
Our calculations for the isotropic Heisenberg model show the emergence of a pronounced minimum
in the differential susceptibility $dM/dB$ at $B_{\text{sat}}/3$ 
as the temperature $T$ is raised from $0$~K for structures based on corner-sharing triangles, specifically
the octahedron, cuboctahedron, and icosidodecahedron.
We also provide the first experimental evidence for this behavior $dM/dB$: It is exhibited by the giant Keplerate magnetic molecule \mofe\ ($\textrm{Fe}^{3+}$ ions with spin $s=5/2$ on the 30 vertices of an icosidodecahedron). The minimum in $dM/dB$ is due to the fact that for low temperatures when $B \approx B_{\text{sat}}/3$ there exist two competing families of spin configurations of which one behaves magnetically ``stiff'' leading to a
reduction of the differential susceptibility. 
\end{abstract}

\maketitle


The magnetism of frustrated one-, two-, and three-dimensional lattice spin
systems is a fascinating subject due to the richness of
phenomena that are observed \cite{Gre:JMC01, Diep94, LhM02}. In this Letter we report that one effect of geometrical frustration, which so far has been reported \cite{Zhi:PRL02} only for the theoretical model of classical spins on a Kagom\'e lattice, already appears for a class of {\it zero-dimensional} materials, namely certain magnetic molecules hosting highly symmetric arrays of classical or quantum spins. These molecular units \cite{Mueller:CCR01} contain a set of paramagnetic ions whose mutual interactions are described by isotropic Heisenberg exchange and where the {\it inter}molecular magnetic interactions (dipole-dipole for the most part) are negligible as compared to {\it intra}molecular 
Heisenberg exchange. Magnetic molecules as zero-dimensional spin systems provide a new 
avenue for detailed exploration of the basic issues of geometric 
frustration. They are particularly appealing since they offer the 
prospect of being modeled unencumbered by some of the complications of 
bulk magnetic materials. 

We here report experimental and theoretical results for the occurrence of a striking anomaly in the  
differential susceptibility $dM/dB$ versus magnetic field $B$ that is
exhibited by the giant Keplerate magnetic molecule \mofe\ \cite{MSS:ACIE99, MLS:CPC01}. This molecule features 30 $\textrm{Fe}^{3+}$ ions 
on the vertices of an icosidodecahedron that interact via nearest-neighbor (nn) isotropic antiferromagnetic (AF) exchange ($J/k_B = 1.57$~K). Due to their near-perfect $O_h$-symmetric coordination environment, the $\textrm{Fe}^{3+}$ ions represent ideal $s=5/2$ spin centers with virually no single-ion anisotropy.
We also present theoretical results for the classical and quantum Heisenberg model showing that the same anomaly in $dM/dB$
occurs for a class of geometrically frustrated zero-dimensional systems, where 
spins mounted on the vertices of a triangle, octahedron, cuboctahedron, or an icosidodecahedron interact via nn isotropic AF exchange. 
As the temperature $T$ is raised from 0 K a deep narrow minimum in $dM/dB$ 
emerges in the vicinity of one-third the saturation field $B_{\text{sat}}$, which upon increasing
$T$ extends 
over a larger field interval and its sharp features progressively deteriorate. 
We attribute this phenomenon to a common topological property
of these polytopes, namely that each is assembled from corner-sharing triangles. 
In the classical case the drop in $dM/dB$ 
can be understood as a result of the interplay of two effects: 
In the immediate vicinity of $B_{\text{sat}}/3$ 
a family of ``up-up-down'' ({\it uud}) spin configurations are energetically competitive with the 
continuous family of spin configurations of lowest energy \cite{Kaw:JPSJ85}. However, the {\it uud} spin configurations 
are magnetically ``stiff'', i.e. $dM/dB \approx 0$ for low temperatures, and thus reduce the susceptibility of the system. 

We write the AF Heisenberg Hamiltonian as
\begin{equation}
\label{Hamiltonian}
H = J \sum_{(m,n)} \tilde{\bm{S}}_m \cdot \tilde{\bm{S}}_n + g \mu_B \bm{B} \cdot \sum_n \tilde{\bm{S}}_n \text{,}
\end{equation}
where $J$ is a positive energy, the spin operators $\tilde{\bm{S}}_n$ are in units of $\hbar$, $\bm{B}$ is the external field, $g$ is the 
spectroscopic splitting factor, $\mu_B$ is the Bohr magneton, and $(m,n)$ directs that the sum is over distinct nearest-neighbor pairs.
The classical counterpart of Eq.(\ref{Hamiltonian}) is obtained by replacing each spin operator $\tilde{\bm{S}}_n$
by $\sqrt{s(s+1)}\bm{S}_n$, where $\bm{S}_n$ is a c-number unit vector \cite{Fis:AJP64, CLA:PRB99}. 

\begin{figure}[!ht]
\begin{center}
\epsfig{file=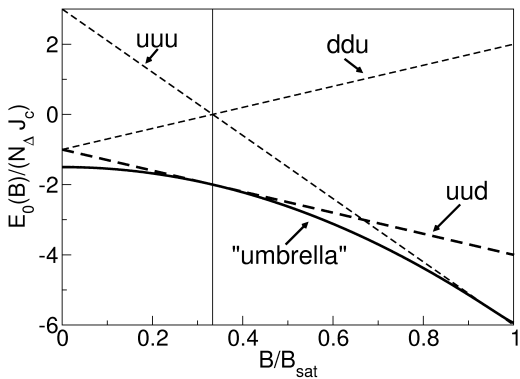,width=68mm}
\vspace*{1mm}
\caption[]{Total energy vs. magnetic field for $T=0$ K for the classical AF triangle, octahedron, cuboctahedron, and icosidodecahedron.
The solid curve is given by Eq.~\ref{groundstate}. The dashed curves
correspond to collinear structures discussed in the text.}
\label{energy}
\end{center}
\end{figure}

One very attractive feature of the polytopes under consideration is that their exact classical ground state energy is 
known \cite{AxL:PRB01}. For $B \le B_{sat}$ it is given by 
\begin{equation}
\label{groundstate}
E_0(B) = - \frac32 N_\Delta J_c [ 1+ 3\left(\frac{B}{B_{sat}}\right)^2 ],
\end{equation}
where $J_c=s(s+1)J$ is called the classical Heisenberg exchange constant, $B_{\text{sat}}= 6 J_c / \mu_c$, $\mu_c = g \mu_B \sqrt{s(s+1)}$, and
$N_\Delta$ is the number of corner-sharing triangles ($= 4, 8, 20$ for the octahedron, cuboctahedron, and icosidodecahedron, respectively).  
A plot of this quantity versus $B/B_{\text{sat}}$ is shown in Fig.~\ref{energy} (solid curve).
The ground state magnetic moment and differential susceptibilty are given by $M_0(B) = -dE_0/dB$ and $dM_0(B)/dB$, respectively. 
For $B = 0$ each spin system is decomposable into 3 sublattices of 
$N/3$ spins each; all spins of a given sublattice are mutually parallel; the sublattices are characterized by three coplanar unit 
vectors with angular spacings of $120^\circ$. The magnetization of the system is linear in $B$ until $B_{\text{sat}}$ and constant
(fully saturated configuration) for larger fields. The linear rise with $B$ can be pictured in terms of the 
folding of an ``umbrella'' \cite{Kaw:JPSJ85} defined by the three sublattice unit vectors as they close towards the field vector $\bm{B}$.

\begin{figure}[!ht]
\begin{center}
\epsfig{file=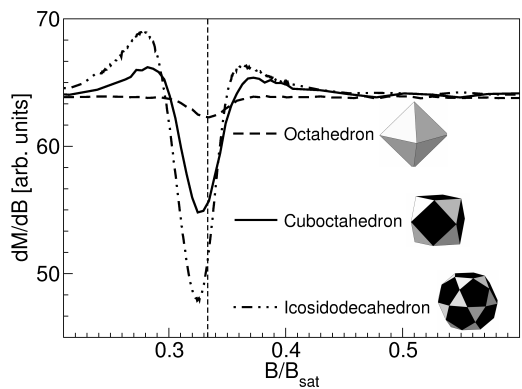,width=68mm}
\vspace*{1mm}
\caption[]{Low-temperature ($k_B T/J_c = 10^{-2}$) simulational results for $dM/dB$ vs. $B$ for classical spins on 
the octahedron, cuboctahedron, and icosidodecahedron.}
\label{all_dmdh}
\end{center}
\end{figure}

\begin{figure}[!ht]
\begin{center}
\epsfig{file=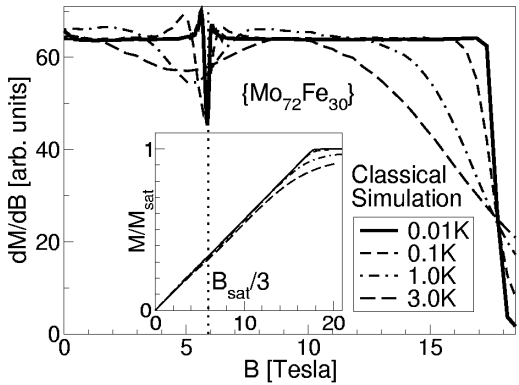,width=68mm}
\vspace*{1mm}
\caption[]{Differential susceptibility $dM/dB$ versus $B$ for the classical Heisenberg model of
\mofe\ obtained by Monte Carlo simulations for temperatures given in the legend.}
\label{Fe30_theory}
\end{center}
\end{figure}

Also shown in Fig.~\ref{energy} are the energy curves for three other specific configurations of interest.
These are configurations where the three (unit) spin vectors associated with each triangle are constrained
to be collinear and the resultant vector is either parallel or anti-parallel to $\bm{B}$. These configurations are labeled as
{\it uuu} (up-up-up), {\it uud}, and {\it ddu}.
For each of these collinear configurations the magnetic moment of the polytope is independent of $B$ and thus $dM/dB$ vanishes and one can describe
these configurations as being magnetically ``stiff''. The fully saturated {\it uuu} configuration is of minimal energy for $B>B_{\text{sat}}$.
The {\it uud} configuration is of special interest since its energy coincides with the minimal 
energy of the spin system for $B=B_{\text{sat}}/3$ and exceeds the minimal energy for any other choice of field.
For $T=0$~K and for any choice of $B$ other than $B_{\text{sat}}$ the {\it uud} configuration will not play a role.
However, for $T>0$~K and for $B$ in the vicinity of $B_{\text{sat}}/3$ a significant contribution to the partition function
will arise from the set of configurations derived by infinitesimal modifications of the {\it uud} configuration.
These slightly modified {\it uud} configurations lead to a reduction of the differential susceptibility of the system because of their
magnetic stiffness.
Our qualitative considerations for $T>0$~K are confirmed by the results of our classical 
Monte Carlo simulations for the three polytopes as shown in Fig.~\ref{all_dmdh}.
Fig.~\ref{Fe30_theory} displays the results for a classical model of \mofe, namely 30 classical spins on the 
vertices of an icosidodecahedron, as substitutes for quantum spins with $s=5/2$.
As $T$ is increased from $0$ K a sharp narrow
drop emerges that is situated at $B_{\text{sat}}/3$ (vertical dashed line). As $T$ 
continues to increase the drop extends over a larger interval and its sharp 
features progressively wash away. One also observes a 
temperature dependence of the field associated with the minimum in $dM/dB$, i.e., it decreases with increasing $T$. 

\begin{figure}[!ht]
\begin{center}
\epsfig{file=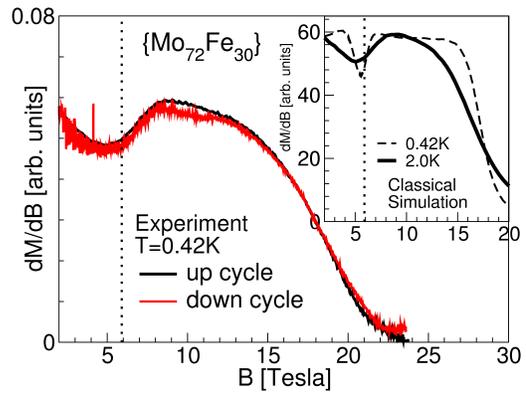,width=68mm}
\vspace*{1mm}
\caption[]{Experimental results (in arbitrary units) for a sample of \mofe\
performed at $0.42$~K using a pulsed-field technique.In the inset Monte Carlo results for $0.42$~K and $2.0$~K are given.}
\label{Fe30_exp}
\end{center}
\end{figure}

The relevance of these theoretical results to real magnetic materials
is demonstrated by our experimental data for the differential susceptibilty of the giant Keplerate magnetic molecule \mofe. The magnetization was measured at $0.42$~K in pulsed magnetic fields up to 23 Tesla (sweep rate 15000 Tesla/s) at the Okayama High Magnetic Field Laboratory by using a standard inductive method. The sample was immersed in liquid $^3$He to maintain good contact with the
thermal bath. The experimental results for $dM/dB$ (in arbitrary units) are shown in Fig.~\ref{Fe30_exp} and the drop at about $B_{\text{sat}}/3$ is clearly evident. 
However, the data resembles the simulational curve for $2$~K (see inset Fig.~\ref{Fe30_exp}), not $0.42$~K, perhaps suggesting an elevated effective spin temperature due to the high sweep rate. To clarify this point we also measured the magnetization in steady fields by a capacitance method in the range up to $7$ Tesla for $T=16$~mK and $0.73$~K and obtained close agreement with the pulsed-field data. The considerable broadening of the drop in $dM/dB$ may point to the occurence of a staggered field \cite{Mats:JPSJ76}, since the principal axes of the Fe ions are not strictly equivalent. A specific suggestion \cite{Has:JPSJ04} is that the broadening is due to Dzyaloshinskii-Moriya terms supplementing the isotropic Heisenberg model, originating from possible low symmetry of the nearest-neighbor Fe-Fe bond. Further study of this issue is warranted.

\begin{figure}[!ht]
\begin{center}
\epsfig{file=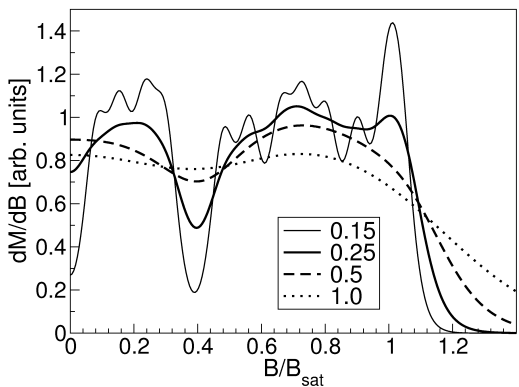,width=68mm}
\vspace*{1mm}
\caption[]{Differential susceptibility $dM/dB$ versus $B/B_{\text{sat}}$ of the quantum Heisenberg
cuboctahedron ($s=1$) for values of $k_B T/J$ shown in the legend.}
\label{cubo_q}
\end{center}
\end{figure}

To explore the role of quantum effects we have calculated $dM/dB$ for the triangle and the octahedron of spins with arbitrary $s$ as well as for a 
cuboctahedron ($N=12$) with $s=1/2$ and $s=1$. For the latter system this involves numerical diagonalization of matrices
defined on a Hilbert space of dimension $3^{12}$ ($= 531441$). Even by fully exploiting the symmetries 
of the Hamiltonian this is at the limit of present day computing capabilities. The results for the cuboctahedron with $s=1$ are shown in Fig.~\ref{cubo_q} for 
different temperatures. As in the previous figures one again encounters a strong reduction of $dM/dB$. The minimum is located exactly at $M_{\text{sat}}/3$, but since $B$ and $M$ are not strictly 
proportional for a quantum system the drop occurs for fields slightly larger than $B_{\text{sat}}/3$.
For $T=0$~K $M$ vs. $B$ can be described as a ``staircase'' of 12 steps originating from ground state Zeeman level crossings and
$dM/dB$ consists of a set of Dirac delta functions at the crossing fields.
For $T>0$~K the abrupt magnetization steps are smoothed out and $dM/dB$ exhibits finite peaks.
Our results for the triangle and the octahedron for general spins $s$ 
exhibit the same overall behavior seen in Fig.~\ref{cubo_q}.   

One can understand that the pronounced minimum in the susceptibility occurs for classical as well quantum spins by examining the 
partition function for the particularly simple example
of the triangle where the results can be obtained by exact analytical methods.
For integer spins $s$ the quantum partition function may be written as
\begin{equation}
\label{z.quantum}
Z(t,b) = \left( \sinh(b \sigma_0) \right)^{-1} \sum_{n=0}^{3s} G_n \, e^{-\frac{\sigma_n^2}{2t}} \sinh(b \sigma_n),
\end{equation}
where $b=\mu_c B/(k_B T)$, $t=k_B T/J_c$, $\sigma_n = (n+\frac12)/\sqrt{s(s+1)}$,
$G_n=\Gamma_n/\sqrt{s(s+1)}$, and $\Gamma_n$ is the multiplicity factor, namely the number
of distinct ways of achieving total spin $n$ upon adding three distinct (integer) quantum spins $s$.
In particular $\Gamma_n = 2(n+\frac12)$ for $0 \leq n \leq s$ and $\Gamma_n = 3(s+\frac12) -
(n+\frac12)$ for $s+1 \leq n \leq 3s$. The analogous formulas are easily derived for half-integer spins $s$.
Formula (\ref{z.quantum}) for $Z$ is very similar to that for the classical Heisenberg triangle which may be written as \cite{CLA:PRB99}
\begin{equation}
\label{z.classical}
Z(t,b) = b^{-1} \int_0^3 \text{d} S \, G(S)\, e^{-\frac{S^2}{2t}} \sinh (bS).
\end{equation}
Here $G(S) = 2S$ for $0 \leq S \leq 1$ and $G(S) = 3-S$ for $1 \leq S \leq 3$, arising from
considering the geometrical volume available to three unit vectors such that the magnitude of their vector sum
lies within a shell of radius S and unit thickness. Indeed it is straightforward to verify that in the limit
$s \to \infty$ the quantum result (Eq.~(\ref{z.quantum})) agrees with the classical formula (Eq.~(\ref{z.classical})). In the quantum formula the
multiplicity factor corresponds to the classical geometrical function $G(S)$. Each of these quantities has two distinct
branches, depending on whether $n$ is in the interval $[0,s]$ or $[s+1,3s]$ or whether $S$ is
in the interval $[0,1]$ or $[1,3]$. In fact, the existence of two distinct branches becomes manifest
in various higher derivatives of $Z(t,b)$ at nonzero temperatures for fields in the vicinity of
$B=B_{\text{sat}}/3$. For $0 < t \ll 1$ there exists a narrow field range at about $B_{\text{sat}}/3$ such
that each of the functions $\exp (-\sigma_n^2/(2t)) \sinh(b
\sigma_n)$ and $\exp (-S^2/(2t)) \sinh(bS)$ has a very narrow maximum for $\sigma_n \approx 1$ and 
$S \approx 1$ but nevertheless samples the two
branches. This is the mathematical orgin of the pronounced minimum in $dM/dB$ at $B=B_{\text{sat}}/3$.   

Plateau like structures in the magnetization versus B in various two- and three-dimensional lattices built of corner-sharing triangles 
lattices at one-third of the saturated moment have been under investigation for the past two decades as an expression of geometric frustration
\cite{Gre:JMC01, Diep94, LhM02, Nar:EPL04, Jac:JPSJ93}. Moreover, theoretical studies of the classical Heisenberg antiferromagnet on the Kagom\'e lattice
show that $dM/dB$ 
has a pronounced minimum at one-third of $B_{\text{sat}}$ \cite{Zhi:PRL02}.
However, the study of selective magnetic molecules such as \mofe\
can give new insights for this subject since such molecules
are much better accessible both experimentally and theoretically.

In summary, we have shown that for a class of geometrically frustrated magnetic polytopes, 
namely the octahedron, the cuboctahedron and the icosidodecahedron, field-induced 
competitive spin configurations exist which manifest themselves in an pronounced minimum
in the differential susceptibility $dM/dB$ in the vicinity of $B_{\text{sat}}/3$.
We have also reported the first experimental observation of this effect. Our data for the giant Keplerate magnetic molecule \mofe\ are consistent with our classical Monte Carlo results for the icosidodecahedron. Furthermore, we have shown that this 
feature reflects a general intrinsic property of the very building block of these specific 
polytopes, namely the simple AF equilateral Heisenberg spin triangle, and emerges for both 
classical and quantum spins. Moreover, our theoretical calculations for each of these polytopes show that the specific heat versus $B$ also exhibits anomalous 
behavior in the vicinity of $B_{\text{sat}}/3$ \cite{Schr04}.
A measurement of this quantity for \mofe\ at very low temperatures would be of great interest. 

\section*{Acknowledgments}
We acknowledge useful discussions with H.-J. Schmidt. H. Nojiri acknowledges the support by Grant in Aid by Scientific Research from MEXT, Japan and by Uesuko Science Foundation.
Ames Laboratory is operated for the U.S. Department of Energy by Iowa State University 
under Contract No. W-7405-Eng-82.

\begin{thebibliography}{15}
\expandafter\ifx\csname natexlab\endcsname\relax\def\natexlab#1{#1}\fi
\expandafter\ifx\csname bibnamefont\endcsname\relax
  \def\bibnamefont#1{#1}\fi
\expandafter\ifx\csname bibfnamefont\endcsname\relax
  \def\bibfnamefont#1{#1}\fi
\expandafter\ifx\csname citenamefont\endcsname\relax
  \def\citenamefont#1{#1}\fi
\expandafter\ifx\csname url\endcsname\relax
  \def\url#1{\texttt{#1}}\fi
\expandafter\ifx\csname urlprefix\endcsname\relax\def\urlprefix{URL }\fi
\providecommand{\bibinfo}[2]{#2}
\providecommand{\eprint}[2][]{\url{#2}}

\bibitem[{\citenamefont{Greedan}(2001)}]{Gre:JMC01}
\bibinfo{author}{\bibfnamefont{J.}~\bibnamefont{Greedan}}, \bibinfo{journal}{J.
  Mater. Chem.} \textbf{\bibinfo{volume}{11}}, \bibinfo{pages}{37}
  (\bibinfo{year}{2001}).

\bibitem[{\citenamefont{Diep}(1994)}]{Diep94}
\bibinfo{editor}{\bibfnamefont{H.}~\bibnamefont{Diep}}, ed.,
  \emph{\bibinfo{title}{Magnetic systems with competing interactions}}
  (\bibinfo{publisher}{World Scientific}, \bibinfo{address}{Singapore},
  \bibinfo{year}{1994}).

\bibitem[{\citenamefont{Lhuillier et~al.}(2002)\citenamefont{Lhuillier, in:
  C.~Berthier, Levy, and Martinez}}]{LhM02}
\bibinfo{editor}{\bibfnamefont{C.}~\bibnamefont{Lhuillier} \bibnamefont{and} \bibfnamefont{G.~Misguich}},
  \bibinfo{editor}{\bibnamefont{in: C.~Berthier}},
  \bibinfo{editor}{\bibfnamefont{L.}~\bibnamefont{Levy}}, \bibnamefont{and}
  \bibinfo{editor}{\bibfnamefont{G.}~\bibnamefont{Martinez}}, eds.,
  \emph{\bibinfo{title}{High Magnetic Fields}} (\bibinfo{publisher}{Springer},
  \bibinfo{address}{Berlin}, \bibinfo{year}{2002}), pp.
  \bibinfo{pages}{161--190}, \bibinfo{note}{cond-mat/0109146}.

\bibitem[{\citenamefont{Zhitomirsky}(2002)}]{Zhi:PRL02}
\bibinfo{author}{\bibfnamefont{M.~E.} \bibnamefont{Zhitomirsky}},
  \bibinfo{journal}{Phys. Rev. Lett.} \textbf{\bibinfo{volume}{88}},
  \bibinfo{pages}{057204} (\bibinfo{year}{2002}).

\bibitem[{\citenamefont{M{\"u}ller et~al.}(2001)\citenamefont{M{\"u}ller,
  K{\"o}gerler, and Dress}}]{Mueller:CCR01}
\bibinfo{author}{\bibfnamefont{A.}~\bibnamefont{M{\"u}ller}},
  \bibinfo{author}{\bibfnamefont{P.}~\bibnamefont{K{\"o}gerler}},
  \bibnamefont{and} \bibinfo{author}{\bibfnamefont{A.~W.~M.}
  \bibnamefont{Dress}}, \bibinfo{journal}{Coord. Chem. Rev.}
  \textbf{\bibinfo{volume}{222}}, \bibinfo{pages}{193} (\bibinfo{year}{2001}).

\bibitem[{\citenamefont{M\"uller et~al.}(1999)\citenamefont{M\"uller, Sarkar,
  Shah, B\"ogge, Schmidtmann, Sarkar, K\"ogerler, Hauptfleisch, Trautwein, and
  Sch\"unemann}}]{MSS:ACIE99}
\bibinfo{author}{\bibfnamefont{A.}~\bibnamefont{M\"uller}},
  \bibinfo{author}{\bibfnamefont{S.}~\bibnamefont{Sarkar}},
  \bibinfo{author}{\bibfnamefont{S.~Q.~N.} \bibnamefont{Shah}},
  \bibinfo{author}{\bibfnamefont{H.}~\bibnamefont{B\"ogge}},
  \bibinfo{author}{\bibfnamefont{M.}~\bibnamefont{Schmidtmann}},
  \bibinfo{author}{\bibfnamefont{S.}~\bibnamefont{Sarkar}},
  \bibinfo{author}{\bibfnamefont{P.}~\bibnamefont{K\"ogerler}},
  \bibinfo{author}{\bibfnamefont{B.}~\bibnamefont{Hauptfleisch}},
  \bibinfo{author}{\bibfnamefont{A.}~\bibnamefont{Trautwein}},
  \bibnamefont{and}
  \bibinfo{author}{\bibfnamefont{V.}~\bibnamefont{Sch\"unemann}},
  \bibinfo{journal}{Angew. Chem., Int. Ed.} \textbf{\bibinfo{volume}{38}},
  \bibinfo{pages}{3238} (\bibinfo{year}{1999}).

\bibitem[{\citenamefont{M\"uller et~al.}(2001)\citenamefont{M\"uller, Luban,
  Schr\"oder, Modler, K\"ogerler, Axenovich, Schnack, Canfield, Bud'ko, and
  Harrison}}]{MLS:CPC01}
\bibinfo{author}{\bibfnamefont{A.}~\bibnamefont{M\"uller}},
  \bibinfo{author}{\bibfnamefont{M.}~\bibnamefont{Luban}},
  \bibinfo{author}{\bibfnamefont{C.}~\bibnamefont{Schr\"oder}},
  \bibinfo{author}{\bibfnamefont{R.}~\bibnamefont{Modler}},
  \bibinfo{author}{\bibfnamefont{P.}~\bibnamefont{K\"ogerler}},
  \bibinfo{author}{\bibfnamefont{M.}~\bibnamefont{Axenovich}},
  \bibinfo{author}{\bibfnamefont{J.}~\bibnamefont{Schnack}},
  \bibinfo{author}{\bibfnamefont{P.~C.} \bibnamefont{Canfield}},
  \bibinfo{author}{\bibfnamefont{S.}~\bibnamefont{Bud'ko}}, \bibnamefont{and}
  \bibinfo{author}{\bibfnamefont{N.}~\bibnamefont{Harrison}},
  \bibinfo{journal}{ChemPhysChem} \textbf{\bibinfo{volume}{2}},
  \bibinfo{pages}{517} (\bibinfo{year}{2001}).

\bibitem[{\citenamefont{Kawamura and Miyashita}(1985)}]{Kaw:JPSJ85}
\bibinfo{author}{\bibfnamefont{H.}~\bibnamefont{Kawamura}} \bibnamefont{and}
  \bibinfo{author}{\bibfnamefont{S.}~\bibnamefont{Miyashita}},
  \bibinfo{journal}{J.~Phys. Soc. Japan} \textbf{\bibinfo{volume}{54}},
  \bibinfo{pages}{4530} (\bibinfo{year}{1985}).

\bibitem[{\citenamefont{Fisher}(1964)}]{Fis:AJP64}
\bibinfo{author}{\bibfnamefont{M.}~\bibnamefont{Fisher}}, \bibinfo{journal}{Am.
  J. Phys.} \textbf{\bibinfo{volume}{32}}, \bibinfo{pages}{343}
  (\bibinfo{year}{1964}).

\bibitem[{\citenamefont{Ciftja et~al.}(1999)\citenamefont{Ciftja, Luban,
  Auslender, and Luscombe}}]{CLA:PRB99}
\bibinfo{author}{\bibfnamefont{O.}~\bibnamefont{Ciftja}},
  \bibinfo{author}{\bibfnamefont{M.}~\bibnamefont{Luban}},
  \bibinfo{author}{\bibfnamefont{M.}~\bibnamefont{Auslender}},
  \bibnamefont{and} \bibinfo{author}{\bibfnamefont{J.~H.}
  \bibnamefont{Luscombe}}, \bibinfo{journal}{Phys. Rev. B}
  \textbf{\bibinfo{volume}{60}}, \bibinfo{pages}{10122} (\bibinfo{year}{1999}).

\bibitem[{\citenamefont{Axenovich and Luban}(2001)}]{AxL:PRB01}
\bibinfo{author}{\bibfnamefont{M.}~\bibnamefont{Axenovich}} \bibnamefont{and}
  \bibinfo{author}{\bibfnamefont{M.}~\bibnamefont{Luban}},
  \bibinfo{journal}{Phys. Rev. B} \textbf{\bibinfo{volume}{63}},
  \bibinfo{pages}{100407} (\bibinfo{year}{2001}).

\bibitem[{\citenamefont{Matsuura and Ajiro}(1976)}]{Mats:JPSJ76}
\bibinfo{author}{\bibfnamefont{M.}~\bibnamefont{Matsuura}} \bibnamefont{and}
  \bibinfo{author}{\bibfnamefont{Y.}~\bibnamefont{Ajiro}},
  \bibinfo{journal}{J. Phys. Soc. Japan} \textbf{\bibinfo{volume}{41}},
  \bibinfo{pages}{44} (\bibinfo{year}{1976}).

\bibitem[{\citenamefont{Hasegawa and Shiba}(2004)}]{Has:JPSJ04}
\bibinfo{author}{\bibfnamefont{M.}~\bibnamefont{Hasegawa}} \bibnamefont{and}
  \bibinfo{author}{\bibfnamefont{H.}~\bibnamefont{Shiba}},
  \bibinfo{journal}{J. Phys. Soc. Japan} \textbf{\bibinfo{volume}{73}},
  \bibinfo{pages}{2543} (\bibinfo{year}{2004}).

\bibitem[{\citenamefont{Narumi et~al.}(2004)\citenamefont{Narumi, Katsumata,
  Honda, Domenge, Sindzingre, Lhuillier, Shimaoka, Kobayashi, and
  Kindo}}]{Nar:EPL04}
\bibinfo{author}{\bibfnamefont{Y.}~\bibnamefont{Narumi}},
  \bibinfo{author}{\bibfnamefont{K.}~\bibnamefont{Katsumata}},
  \bibinfo{author}{\bibfnamefont{Z.}~\bibnamefont{Honda}},
  \bibinfo{author}{\bibfnamefont{J.-C.} \bibnamefont{Domenge}},
  \bibinfo{author}{\bibfnamefont{P.}~\bibnamefont{Sindzingre}},
  \bibinfo{author}{\bibfnamefont{C.}~\bibnamefont{Lhuillier}},
  \bibinfo{author}{\bibfnamefont{Y.}~\bibnamefont{Shimaoka}},
  \bibinfo{author}{\bibfnamefont{T.~C.} \bibnamefont{Kobayashi}},
  \bibnamefont{and} \bibinfo{author}{\bibfnamefont{K.}~\bibnamefont{Kindo}},
  \bibinfo{journal}{Europhys. Lett.} \textbf{\bibinfo{volume}{65}},
  \bibinfo{pages}{705} (\bibinfo{year}{2004}).

\bibitem[{\citenamefont{Jacobs et~al.}(1993)\citenamefont{Jacobs, Nikuni, and
  Shiba}}]{Jac:JPSJ93}
\bibinfo{author}{\bibfnamefont{A.~E.} \bibnamefont{Jacobs}},
  \bibinfo{author}{\bibfnamefont{T.}~\bibnamefont{Nikuni}}, \bibnamefont{and}
  \bibinfo{author}{\bibfnamefont{H.}~\bibnamefont{Shiba}}, \bibinfo{journal}{J.
  Phys. Soc. Japan} \textbf{\bibinfo{volume}{62}}, \bibinfo{pages}{4066}
  (\bibinfo{year}{1993}).

\bibitem[{\citenamefont{Schr{\"o}der and Luban}(2004)}]{Schr04}
\bibinfo{author}{\bibfnamefont{C.}~\bibnamefont{Schr{\"o}der}}
  \bibnamefont{and} \bibinfo{author}{\bibfnamefont{M.}~\bibnamefont{Luban}}
  (\bibinfo{year}{2004}), \bibinfo{note}{unpublished}.

\end{thebibliography}

\end{document}